\documentclass[aps,prc,showpacs,twocolumn,superscriptaddress]{revtex4-1}
\usepackage{amsmath}
\usepackage{graphicx}
\usepackage{epsfig,ulem}
\def\beq{\begin{equation}}
\def\eeq{\end{equation}}
\def\beqa{\begin{eqnarray}}
\def\eeqa{\end{eqnarray}}
\def\ban{\begin{eqnarray*}}
\def\ean{\end{eqnarray*}}
\def\bi{\begin{itemize}}
\def\ei{\end{itemize}}

\begin{document}
\title{Compact stars within an asy-soft quark-meson-coupling model}
\author{Prafulla K. Panda}
\affiliation{Centro de F\'{\i}sica Computacional - Departamento de F\'{\i}sica,
Universidade de Coimbra - P-3004 - 516 - Coimbra - Portugal}
\affiliation{Department of Physics, C.V. Raman College of Engeneering,
Vidya Nagar, Bhubaneswar-752054, India}
\author{Alexandre M. S. Santos}
\affiliation{Centro de F\'{\i}sica Computacional - Departamento de F\'{\i}sica,
Universidade de Coimbra - P-3004 - 516 - Coimbra - Portugal}
\affiliation{Universidade Federal de Santa Catarina - Campus Curitibanos,
Caixa Postal 101 - 89.520-000 - Curitibanos-SC - Brazil}
\author{D\'ebora P. Menezes}
\affiliation{Departamento de F\'isica, CFM, Universidade Federal de Santa 
Catarina, CP 476, CEP 88.040-900 Florian\'opolis - SC - Brazil}
\author{Constan\c ca Provid\^encia}
\affiliation{Centro de F\'{\i}sica Computacional - Departamento de F\'{\i}sica,
Universidade de Coimbra - P-3004 - 516 - Coimbra - Portugal}
\begin{abstract}
We investigate compact star properties within the quark meson coupling model 
(QMC) with a soft symmetry energy density dependence at large densities. In 
particular,  the hyperon content and the mass/radius curves for the families 
of stars obtained within the model are discussed. The hyperon-meson couplings 
are chosen according to experimental values of the hyperon nuclear matter 
potentials, and possible uncertainties are considered. It is shown that
a softer symmetry energy gives rise to stars with less hyperons, smaller 
radii and larger masses. Hyperon-meson couplings may also have a strong 
effect on the mass of the star.
\end{abstract}
\maketitle
\section{Introduction}

In the last years important efforts have been done to determine the
 density dependence of the symmetry energy of asymmetric nuclear matter
(see the reviews \cite{Baran-05,Steiner-05,Bao-An-08} and references therein).
Correlations between different quantities in bulk matter and
finite nuclei have been established. For instance, the correlation between
the slope of the pressure of neutron matter at $\rho=0.1$ fm$^{-3}$ and the
neutron skin thickness of $^{208}$Pb \cite{BrownB-00,Typel-01}, or the
correlation between the crust-core transition density and the  neutron
skin thickness of $^{208}$Pb \cite{Horowitz-01} are well determined. 
Presently, there also exist different experimental measurements that 
constrain the saturation properties of the symmetry energy \cite{exp}.

The quark-meson-coupling (QMC) model ~\cite{Guichon-88,qmc,panda04} is an effective 
nuclear model that takes into account the internal
structure of the nucleon  explicitly. Within the
QMC model, matter at low densities and temperatures is a
system of nucleons interacting through meson fields, with
quarks and gluons confined within MIT bags \cite{Chodos-74}. For matter
at very high density or temperature, one expects that baryons
and mesons dissolve and that the entire system of quarks and
gluons becomes confined within a single, big, MIT bag. Within QMC it is
possible to describe in a consistent way both nucleons and hyperons \cite{octet}.
The energy of the baryonic MIT bag is identified with the mass of the baryon and
is obtained self-consistently from the calculation. It is important to stress
that within the QMC model the coupling of hyperons to the $\sigma$-meson is
fixed at the level of the saturation properties of the equation of state
(EOS). Hypernuclei properties \cite{schaffner00,lambda,cascade,gal2010} will 
then allow the determination of the coupling of hyperons to
the isoscalar-vector meson without any ambiguity except for the uncertainty on
the experimental hypernuclei data. {Within the non-linear Walecka models (NLWM)
this is not possible and some other constraint must be imposed, such as using
the SU(6) symmetry to fix the hyperon-vector meson couplings \cite{gal84}.}

Recently, new data on neutron stars have been obtained
\cite{demorest,steiner10} that theory should
explain, namely,  the large value  1.97 $\pm$ 0.04 $M_\odot$ of the recent 
mass measurement of the binary millisecond pulsar PSR J16142230 
\cite{demorest}, and the  empirical EOS
obtained by Steiner {\it et al.} from a heterogeneous set of seven
neutron stars with well-determined distances \cite{steiner10} which
predicts quite small radii. These last results,
however, should still be considered with care because there are many
uncertainties involved.

The symmetry energy at saturation is quite well established, however, the
density dependence of the symmetry energy is not so well known and different
models predict a wide range of values for the symmetry energy slope at
saturation. Although the symmetry energy slope of QMC at saturation (94 MeV
\cite{Santos-09}) is within the range of values compatible with experimental
observations \cite{chen05}, most of the experimental observables that 
have been proposed to obtain a measure of the symmetry energy slope, predict
smaller slopes \cite{exp}, which could be as low as 30 MeV. Moreover, the large
value of the QMC slope %will not allow 
{ prohibits} the prediction of small radii as the ones
indicated by the   empirical EOS \cite{steiner10}.

In  \cite{Santos-09} we have
introduced the $\delta$-meson in the QMC model and have studied its  effect on
the density dependence of the symmetry energy and subsaturation instabilities
of nuclear matter. However, the introduction of the $\delta$-meson gives rise
to a stiffer symmetry energy and  this mechanism will not allow us to obtain
 a softer symmetry energy for the QMC.

In the present study we consider an extension of the QMC that includes
a 
%meson 
nonlinear term involving the  $\omega$ and $\rho$ mesons. 
This  term affects the isovector channel of the QMC equation of state,
namely the density dependence of the symmetry energy
\cite{Horowitz-01,Pais-10}, and  choosing the coupling constant 
adequately it is possible to correct the stiff behavior of the symmetry
energy at large densities in the QMC model.

Stellar matter within the modified QMC model will be studied.
We expect that smaller values of the symmetry energy slope will give rise to
smaller star radii: this has been shown both for nucleonic stars 
\cite{Horowitz-01,fatto2010} and for hyperonic stars \cite{ rafael11}. 
In particular, we want to investigate how the  hyperon content of a compact
star is affected by the density dependence of the symmetry energy.
 This was recently done within the NLWM
including  the nonlinear $\omega\rho$ term  \cite{rafael11} and it was shown 
that the hyperon content could be affected by the density dependence of
the symmetry energy, and that the radius of stars with a mass 1-1.4 $M_\odot$ 
increases linearly with the slope, while the mass is not affected.
 However, the  QMC model gives rise to a softer EOS and
generally predicts smaller hyperon fractions in stellar matter. This could
affect the behavior of the star properties and its relation with the 
slope of the symmetry energy.

The paper is organized as follows: in section II an extension of the QMC model
to include the $\omega-\rho$ coupling is discussed, in section III results are
presented and discussed and %some 
{ the final} conclusions are drawn in the last section.

\section{The quark-meson coupling model}
In what follows we present a review of the QMC model and its generalization
to include the isoscalar-isovector $\omega-\rho$ coupling.

In the QMC model, the nucleon in nuclear medium is assumed to be a
static spherical MIT bag in which quarks interact with the scalar ($\sigma$)
and vector ($\omega$, $\rho$) fields, and those
are treated as classical fields in the mean field
approximation (MFA) \cite{Guichon-88,qmc}.
The quark field, $\psi_{q_{i}}$, inside the bag then
satisfies the equation of motion:
\begin{equation}
\Big[i\,\rlap{/}\partial-m_q^*-\gamma^0(g_\omega^q\,
\omega_0+ g^q_\rho t_{3q} b_{03})\Big]
\,\psi_{q_{i}}=0\ ,% 
\label{eq-motion}
\end{equation}
where $q=u,d,s$, $m_q^*=m_q^0-g_\sigma^q\, \sigma$ with $m_q^0$ the current quark mass, and $g_\sigma^q$,
$g_\omega^q$ and $g_\rho^q$
denote the quark-meson coupling constants. 
The energy of a static bag describing baryon $i$ consisting of three 
quarks in ground state is expressed as
\begin{equation}
E^{\rm bag}_i=\sum_q n_q \, \frac{\Omega_{q_{i}}}{R_i}-\frac{Z_i}{R_i}
+\frac{4}{3}\,  \pi \, R_i^3\,  B_N\ ,
\label{ebag}
\end{equation}
where $\Omega_{q_{i}}\equiv \sqrt{x_{q_{i}}^2+(R_i\, m_q^*)^2}$,
 $R_i$ is the
bag radius of baryon $i$, $x_{q_{i}}$ is the dimensionless quark momentum,
 $Z_i$ is a parameter which accounts for zero-point motion
of baryon $i$ and $B_N$ is the bag constant.
The effective mass of a baryon bag at rest
is taken to be $M_i^*=E_i^{\rm bag}.$
The equilibrium condition for the bag is obtained by
minimizing the effective mass, $M_i^*$ with respect to the bag radius
\begin{equation}
\frac{d\, M_i^*}{d\, R_i^*} = 0.
\label{balance}
\end{equation}
We have considered $B_N^{1/4}=211.30306$ MeV and $R_i=0.6$
fm. The unknowns $Z_i$ are given in \cite{panda04}.

\subsection{QMC with coupled $\omega-\rho$ fields}

The consideration of a coupling between the isoscalar and isovector
fields is carried out much like in the manner it was performed in
\cite{Horowitz-01}.
A note is however on demand: we start out from quarks, which find
themselves confined
in a bag \cite{Chodos-74}, and the boundary conditions for achieving confinement hold the same.
The relevant changes (of couplings) and fittings (to the symmetry energy)
are done otherwise for hadronic matter.
The total energy density of the nuclear matter then reads
\begin{eqnarray}
\varepsilon &=& \frac{1}{2}m_\sigma^2 \sigma^2
- \frac{1}{2} m_\omega^2 \omega^2_0
- \frac{1}{2} m_\rho^2 b_{03}^2
- g_\omega^2 g_\rho^2 \Lambda_v b_{03}^2 \omega_0^2\nonumber\\
&+&g_\omega\omega_0 \sum_B x_{\omega B} \rho_B
+ g_\rho b_{03} \sum_B x_{\rho B} t_{3B}  \rho_3\nonumber\\
&+&\sum_B \frac{2J_B+1}{2\pi^2} \int_0^{k_{FB}} k^2 dk
\left[k^2 + M_B^{* 2}\right]^{1/2},
\end{eqnarray}
where $t_{3B}$ is the isospin projection of baryon $B$. For the nucleons we
take  $ x_{\omega B}=x_{\rho B}=1$. The corresponding coefficients for the
hyperons will be discussed later.
In the above expression for the energy density, we have introduced the 
$\omega-\rho$ couplings. 
The chemical potentials, necessary to define the $\beta$-equilibrium conditions, are given by
$$\mu_B=\sqrt{k_{FB}^2+{M^*_B}^2}+g_\omega \omega_0+{g_\rho} t_{3B}
b_{03}.$$

In the above expressions the mean fields for mesons are  determined  by the equations
\begin{eqnarray*}
\frac{\partial \varepsilon}{\partial \sigma}&=&0, \,
\omega_0=\frac{g_\omega}{{m_\omega^*}^2}\sum_B x_{\omega B} \rho_B , \,
b_{03}=\frac{g_\rho}{{m_\rho^*}^2} \sum_B x_{\rho B} t_{3B}  \rho_3,
\label{sig}
\end{eqnarray*}
where ${m_\omega^*}^2={m_\omega^2}-2\Lambda_v g_\rho^2g_\omega^2b_{03}^2$ and
${m_\rho^*}^2={m_\rho^2}-2\Lambda_v g_\rho^2g_\omega^2\omega_0^2$, and  $g_\omega=3 g_\omega^q$ and $g_\rho= g_\rho^q$.

\begin{table}[t]
\centering
\caption{Nuclear matter properties of the models used in the present work. All
quantities are taken at saturation, where $B/A$=15.7 MeV and the compressibility
$K_0=290$ MeV. $\rho_t$ and $Y_{pt}$ are the density and proton
fraction at the crust-core
transition estimated from the thermodynamical spinodal section. 
%{\bf We could have less decimal numbers for the last two columns to
%  have everything in accordance with NL3 and TW.}
}
\label{tab:properties}

  \begin{tabular}{lcclccccccc}
\hline
    Model   &    $ \Lambda_v$                    & $g_\rho$     & $\mathcal E_{sym}$ & $L$   &$\rho_t$&$Y_{p,t}$ \\
                                  &   &             & (MeV)              &
            (MeV) &(fm$^{-3}$) &    \\
\hline
QMC              &0                    & 8.8606      & 33.70              &93.59  &0.076&0.022 \\
QMC$\omega\rho$ & $0.01$ & 8.9837      & 33.02              &84.99   & 0.079 & 0.024 \\
&$0.02$       & 9.1122      & 32.43              &77.35
  & 0.081& 0.025\\
&$0.03$       & 9.2463      & 31.88              &70.55  & 0.083 &0.026\\
&$0.05$       & 9.5335      & 30.87              &59.03   & 0.089&0.029\\
&$0.10$       & 10.3869     & 27.78              &39.04
  & 0.098& 0.033\\
NL3 \cite{Lalazissis-97}&0&8.9480  &37.34&118.30 &0.065 & 0.021\\
TW\cite{tw}&0& 7.3220    &32.76&55.30 &0.084 &0.038 \\
\hline
\end{tabular}
\end{table}
The model parameters are obtained  by fitting the nucleon mass and enforcing
the stability condition for the bag in free space. The desired values of $E_N \equiv   \epsilon/\rho - M = -15.7$~MeV at saturation
$\rho=\rho_0=0.15$~fm$^{-3}$, are achieved by setting $g_\sigma^q=5.981$,
$g_{\omega}=8.954$. We take the standard values for the meson masses, namely $m_\sigma=550$ MeV,
$m_\omega=783$ MeV, and $m_\rho=770$ MeV.
The couplings
$g_\rho$ and $\Lambda_v$ are determined so that ${\cal E}_{sym}=23.27$ MeV at $\rho=0.1$
fm$^{-3}$ ($k_F\sim 1.14$ fm$^{-1}$).  The parameters  $g_{\rho }$ and $\Lambda_v$ are listed in
Table \ref{tab:properties}. 

\section{Results and discussions}

\begin{figure}[!]
\begin{tabular}{l}
\includegraphics[width=0.9\linewidth,angle=0]{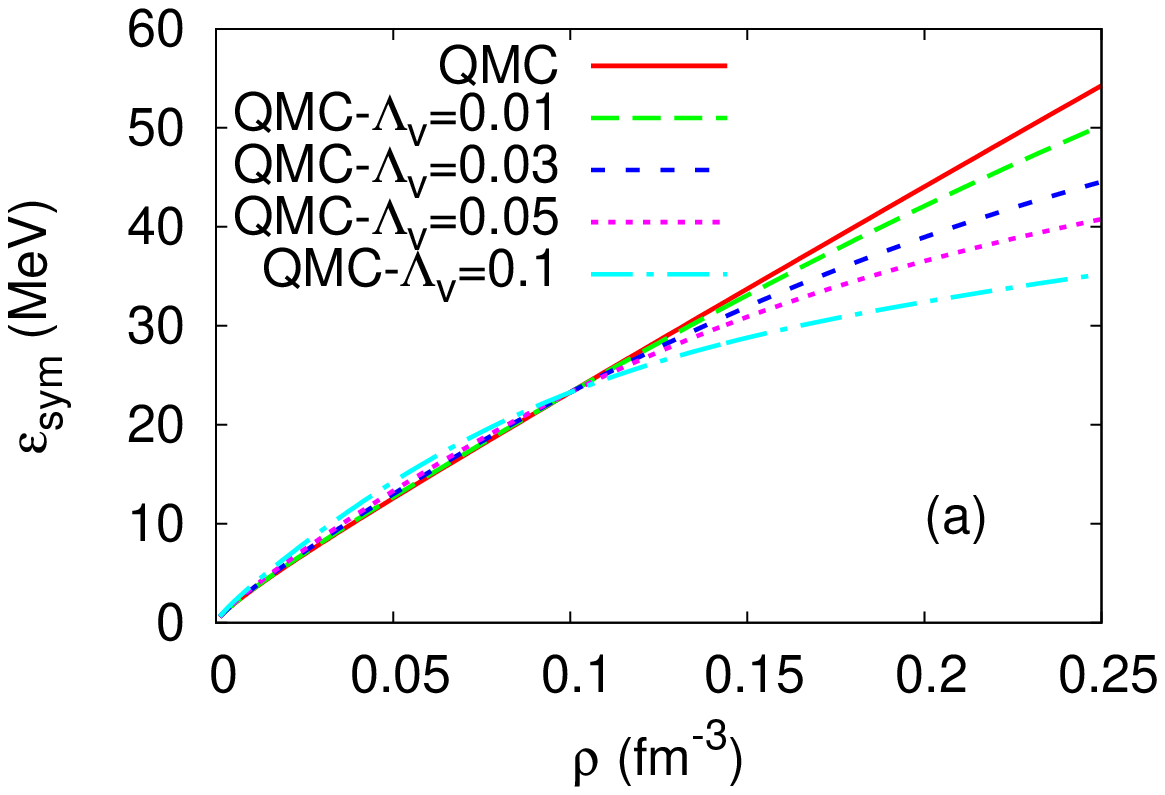}\\
\includegraphics[width=0.9\linewidth,angle=0]{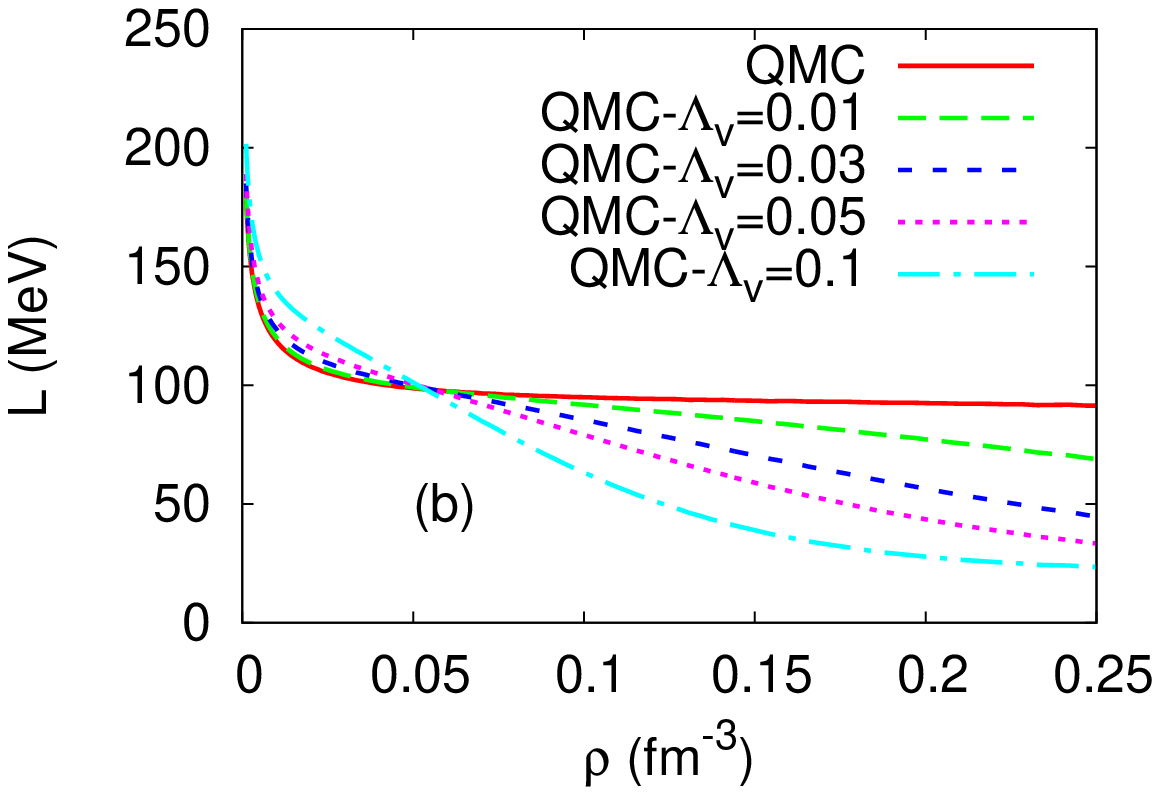}\\
\end{tabular}
\caption{(Color online) Symmetry energy (a),  its slope parameter
$L=3\rho_0 {\cal E}^\prime_{sym}$ (b), and curvature $K_{sym}$ 
for QMC and QMC$\omega\rho$ for different values of the
coupling $\Lambda_v$.
}
\label{fig2}
\end{figure}
Before applying the modified QMC model to the study of stellar matter we
discuss its
properties at saturation and subsaturation densities.

The saturation properties of nuclear matter 
for all the QMC models considered in the present work are shown in Table
\ref{tab:properties}. For comparison, we also include the properties of two well 
known relativistic nuclear models NL3 \cite{Lalazissis-97} and TW \cite{tw}. 
Except for the the symmetry energy of the QMC$\omega\rho$ model with 
$\Lambda_v=0.1$, which could be a bit too low, both the symmetry energy and its
 slope $L=3\rho_0 \partial{\cal E}_{sym}/\partial\rho$  of all the parametrization are well within the experimental constraints coming from different 
sources \cite{exp}.

 Fig.\ref{fig2} shows the dependence on the density of the symmetry energy ${\mathcal E}_{sym}$ and its slope 
for the different parametrizations of the non-linear $\omega-\rho$ term. 
 The symmetry energy  is given by
\begin{equation}
{\mathcal E}_{sym}=\frac{{k_F}^2}{6{\epsilon_F}^2}+
\frac{\rho}{8}\frac{g_\rho^2}{{m_\rho^*}^2}
\label{esym}
\end{equation}
where $\epsilon_F=\sqrt{k_F^2+M_{0}^{*2}}$ and $M_{0}^*$ is the nucleon effective
mass in symmetric nuclear matter. The symmetry energy within QMC shows a rather linear behavior with
density. This is a feature of many NLWM models. The nonlinear $\omega\rho$
term changes the density dependence of the symmetry energy, and, as a result, 
the symmetry energy of the
QMC$\omega\rho$ (Fig.\ref{fig2}(a)) becomes softer at higher densities. This
is confirmed by the 
 slope parameter $L$, plotted in Fig. \ref{fig2}(b): above $\sim 0.06$
 fm$^{-3}$,  $L$  becomes
 smaller  the larger the coupling $\Lambda_v$ is. 
  However, below $\rho=0.1$
fm$^{-3}$ the symmetry energy is larger in the models with a larger
$\Lambda_v$ and this has important effects on the properties of the
crust-core transition.

\subsection{Crust-core transition}

The QMC$\omega\rho$ presents larger
instability regions than QMC at subsaturation densities and  large isospin asymmetries. The larger the
magnitude of the coupling $\Lambda_v$ the smaller the slope $L$ and the larger
the instability region. The same behavior was  obtained in \cite{Pais-10} for
NL3$\omega\rho$ and is contrary to the one obtained in \cite{Santos-09} with the
inclusion of the $\delta$-meson in the QMC model.
In \cite{camille11} the effect of the slope $L$ on the
spinodal surface at large asymmetries was discussed and it was shown that
larger values of $L$ give rise to smaller spinodal regions at large
asymmetries, where matter is closer to neutron matter. Neutron matter pressure
is essentially proportional to the slope $L$ and, therefore, a larger $L$
corresponds to a harder EOS. The crust-core transition densities
are shown in Table I and it is seen that they increase with the increase 
of the $\omega-\rho$ coupling. {The proton fraction at the
  transition density also tends to increase}
%, except for a small  variation around  $\Lambda_v=0.03$.}
QMC gives a  higher transition density than the one
obtained within NL3. For $\Lambda_v$ in the range 0.03-0.05 we get results
similar to TW, as expected due to the values of $L$. 

\begin{figure}[hb!]
\begin{tabular}{lll}
\includegraphics[width=0.9\linewidth,angle=0]{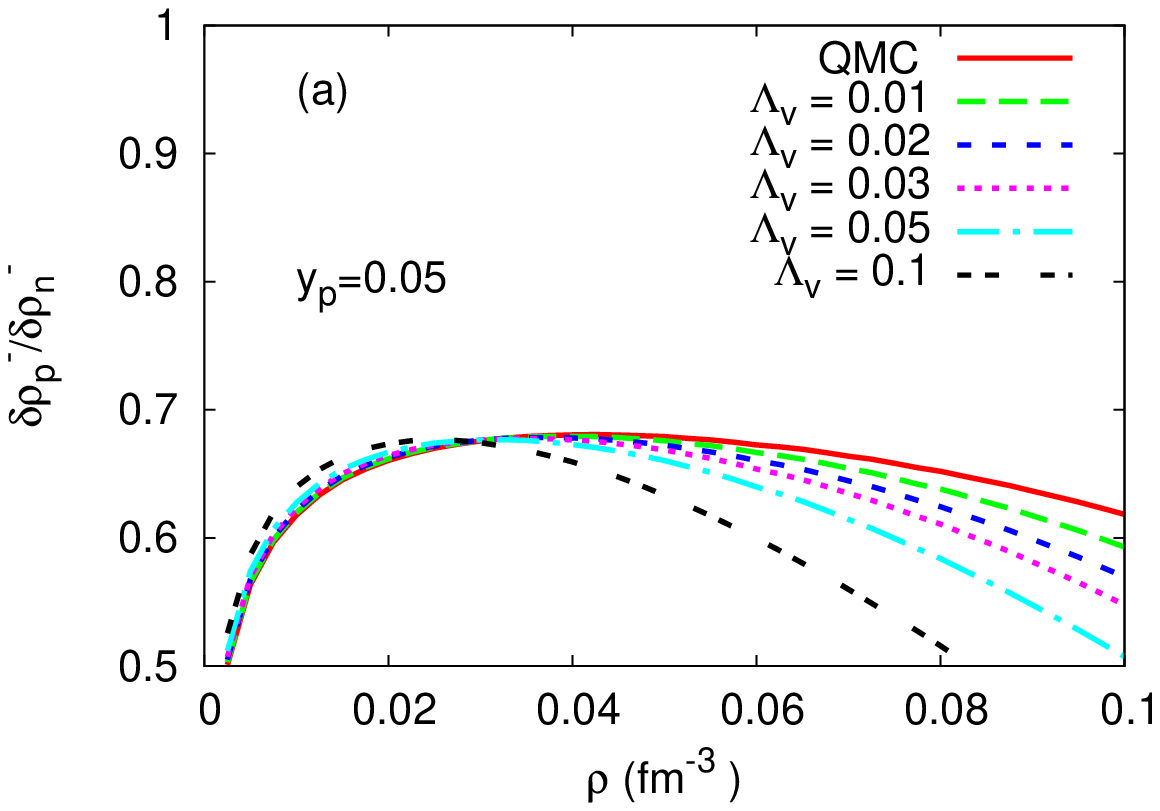}\\
\includegraphics[width=0.9\linewidth,angle=0]{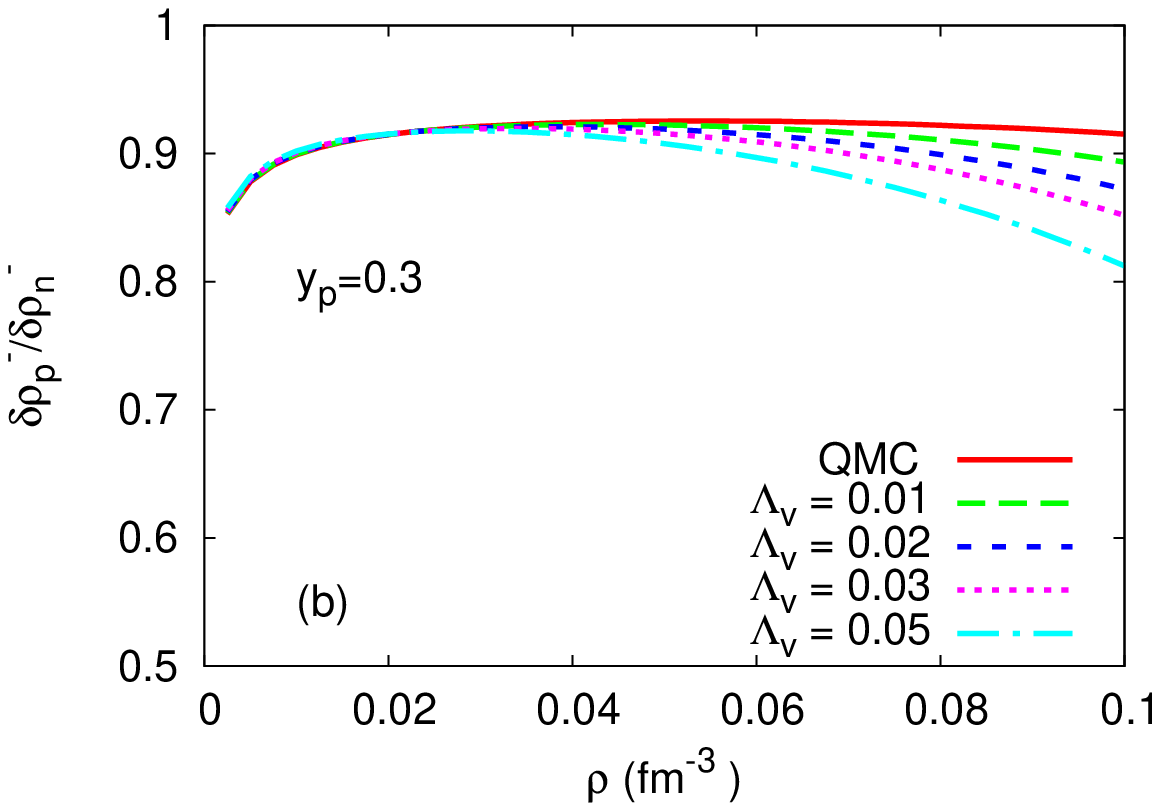}
\end{tabular}
\caption{(Color online) Direction of instability 
for $Y_p$=0.05 (a), %0.1 (b) 
and 0.3 (b). }
\label{fig6}
\end{figure}
In the inner crust of a compact star, matter is not homogeneous,
and is characterized by different 
isospin contents for each phase, i.e.,
the clusterized regions are more isospin symmetric than the surrounding
nuclear gas, the so-called isospin distillation \cite{Xu-08,Chomaz-01-274}.
The extension of the distillation effect
is model dependent and it has been shown that NL3 and other NLWM
parameterizations lead to larger distillation
effects than the density dependent hadron models
\cite{Avancini-06,ProvidenciaC-06,Ducoin-08}.
In Fig.\ref{fig6} we show the ratio
of the proton versus the neutron density fluctuations corresponding to the
unstable mode. This ratio defines the direction of the instability of the
system. We show the results for 
two proton fractions
$Y_p=0.3,$  and $0.05$, for the sake of studying
the effectiveness of the models in restoring the symmetry in the
liquid phase in two situations of interest for compact stars:
$\beta$-equilibrium matter with and without trapped neutrinos.
The $\omega-\rho$
coupling  decreases the $\delta \rho_p/\delta \rho_n$ ratio, as already obtained in \cite{Pais-10} for NL3$\omega\rho$, although in this
last work a dynamical calculation was performed.
Comparing with the result reported in \cite{Avancini-06} we conclude
that: a) QMC behaves differently from NLWM models such as NL3. For
these models the  ratio of the proton versus the neutron density
fluctuations increases with the density, while for QMC after a maximum
obtained at $\rho\sim 0.02$ fm$^{-3}$, this ratio decreases and more
strongly if $\Lambda_v$ is large; b) QMC presents a behavior similar to the
one of relativistic models with density dependent couplings such as TW,
however, the decrease of the distillation effect with density is not so
strong \cite{Avancini-06,Isaac-08}, even for the largest $\Lambda_v$
coupling we have considered.

\subsection{Neutron stars}

Having discussed the behavior of the generalized QMC model at subsaturation
densities, we now turn to the main topic of the present work and discuss the
stellar properties obtained in the present approach.
The composition of the stellar matter is determined by the requirements of the
charge neutrality and chemical equilibrium under the weak processes
\begin{equation}
B_1\rightarrow B_2+l+\bar\nu_l;\hspace{0.5in}B_2+l\rightarrow B_1+\nu_l
\end{equation}
where $B_1$ and $B_2$ are baryons, $l$ is a lepton.
The EOS which depends on the chemical potentials is now modified
according to \cite{octet},  so that the lowest eight baryons are
taken into account. As we restrict ourselves to zero temperature, no
trapped neutrinos are considered, but the electrons and muons are
considered so that charge neutrality and $\beta$-equilibrium can be enforced.
The hyperon couplings are not relevant to the ground state properties of
nuclear matter, but information about them can be available from the levels
in hypernuclei
\cite{chrien,moszk,glen,schaffner00,schaffner02,chiapparini}.
Note that the $s$-quark is unaffected by the sigma and omega
mesons i.e. $g_\sigma^s=g_\omega^s=0\ .$

In QMC the couplings of the hyperons to the $\sigma$-meson do not need
to be fixed because the effective masses of the hyperons are determined
self-consistently at the bag level. Only the  $x_{\omega B}$ and $x_{\rho B}$
have to be fixed. The coupling strength of the $\rho$ meson is given by the
isospin of the baryon, and 
we  obtain $x_{\omega B}$  from the hyperon potentials in nuclear
matter, $U_B=-(M^*_B-M_B) + x_{\omega B}g_{\omega }\omega_0$, for
$B=\Lambda$, $\Sigma$ and $\Xi$ to be
-28 MeV, 30 MeV and -18 MeV, respectively. We find that
$x_{\omega \Lambda}=0.743$, $x_{\omega \Sigma}=1.04$ and $x_{\omega \Xi}=0.346$.
$x_{\rho B}=1$ is fixed for all the baryons. However, while the binding of the
$\Lambda$ to symmetric nuclear matter is well settled experimentally
\cite{lambda}, the  binding values of the $\Sigma^-$ and $\Xi^-$
still have a lot of uncertainties \cite{gal2010}. We, therefore, test the
effect of the coupling to the cascade and show results also for
$V_\Xi=-10$ and 0 MeV.
In fact, measurements from the production of $\Xi$  in
the $^{12}$C$(K^-, K^+)^{12}_{\Xi}$Be are compatible with a shallow attractive
potential $V_{\Xi}\sim -14$ MeV \cite{cascade}.
We obtain  $x_{\omega \Xi}=0.3989$ for
$V_\Xi=-10$ MeV and $x_{\omega \Xi}=0.4643$ for $V_\Xi=0$ MeV.

\begin{figure}[!]
\begin{tabular}{lll}
\includegraphics[width=0.85\linewidth,angle=0]{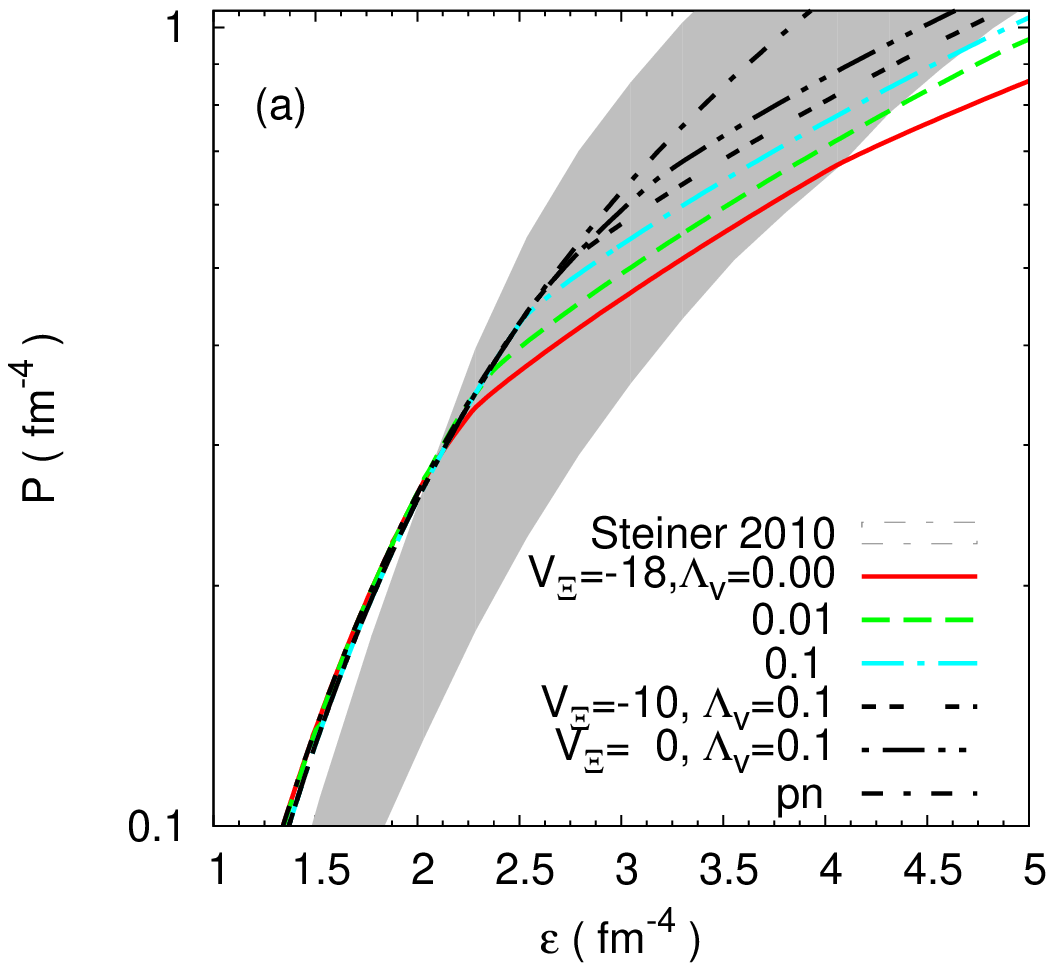}\\
\includegraphics[width=0.8\linewidth,angle=0]{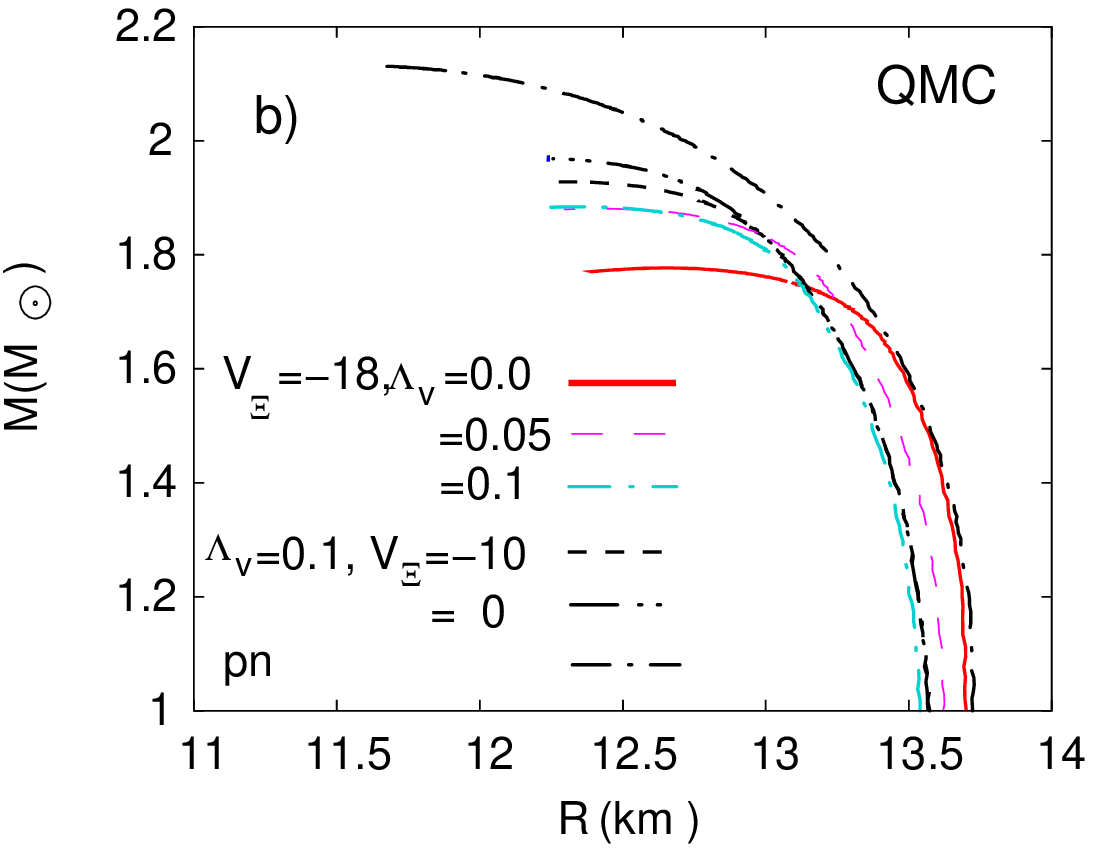}
\end{tabular}
\caption{(Color online) a) Equation of state and b) TOV results for QMC
and QMC$\omega \rho$. The EOS for $V_\Xi=-10$ MeV and 0 MeV where only obtained with
$\Lambda_v=0.1$.
}
\label{fig_8}
\end{figure}

\begin{figure}[!]
\begin{tabular}{lll}
\includegraphics[width=0.85\linewidth,angle=0]{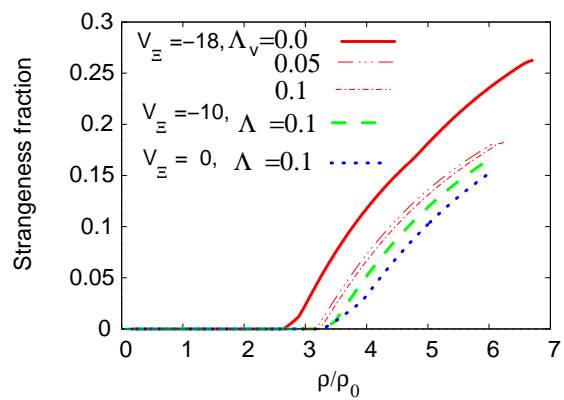}
\end{tabular}
\caption{(Color online) Strangeness fraction as a function of density
 QMC$\omega \rho$. For $V_\Xi=-10$ and 0 MeV we have taken $\Lambda_v=0.1$,
 For $V_\Xi=-18$ MeV, we show results for $\Lambda_v=0,\, 0.05,\, 0.1$. }
\label{fig_6}
\end{figure}

The resulting EOS are displayed in Fig.\ref{fig_8}a) for the QMC,
QMC$\omega \rho$ for different values of the coupling parameter and QMC
with protons and neutrons only. We also include the  empirical EOS
obtained by Steiner {\it et al.} from a heterogeneous set of seven
neutron stars with well-determined distances \cite{steiner10}. 
We conclude that the agreement of the theoretical EOS with 
the empirical one  when hyperons are included in the calculation  is defined by the
hyperon-meson interaction  and the $\Lambda_v$ coupling, or, equivalently,
by the symmetry energy. The QMC $pn$ EOS agrees with the
constraints. {However, the inclusion of hyperons with the hyperon
couplings obtained for the hyperon nuclear potentials taking $V_\Lambda=-28$
MeV, $V_\Sigma=30$ MeV and $V_\Xi=-18$  MeV makes the EOS too soft.}
Increasing $\Lambda_v$ makes the EOS harder
bringing the EOS closer to the constraints defined by the empirical
EOS. This is easily understood with the help of
Fig.\ref{fig_6}. Increasing  $\Lambda_v$ gives rise to a softer $pn$
EOS at high densities and, therefore, hinders the onset of hyperons. So the
larger  $\Lambda_v$ the smaller the hyperon fraction in the star and the
harder the EOS.

The effect of a less attractive $V_\Xi$ potential is also clear: the
EOS becomes harder because the onset of hyperons occurs at larger densities
as shown in Fig.\ref{fig_6}. We conclude that any mechanism than
hinders the formation of hyperons makes the EOS harder.

The EOS enters as input to the Tolman-Volkoff-Oppenheimer \cite{tov}
equations, which generate the macroscopic stellar quantities.
The obtained mass/radius curve for stars with a mass larger than $1M_\odot$
and  the corresponding properties of  maximum mass stars are then shown,
respectively,  in Fig.\ref{fig_8}b) and Table \ref{star}.
\begin{table}
\caption{Stellar properties obtained with the QMC model and different values
of the parameter $\Lambda_v$ and the $\Xi$-meson coupling. $pn$ stands for
nucleonic matter with no hyperons included.}
\begin{tabular}{lccccc}
\hline
$\Lambda_v$& $V_\Xi$(MeV) & $M_{max}(M_{\odot})$ & $M_b(M_{\odot})$ & R(km) &
$\varepsilon_0$(fm$^{-4}$) \\
\hline
0.0 &-18& 1.776 & 2.006 & 12.657 & 4.620 \\
0.01&-18 & 1.836 & 2.096 & 12.496 & 4.837 \\
0.03&-18 & 1.871 & 2.152 & 12.458 & 4.892 \\
0.05&-18 & 1.880 & 2.170 & 12.415 & 4.945 \\
0.1&-18 & 1.888 & 2.182 & 12.345 & 5.004 \\
0.1& -10  & 1.928 & 2.243 & 12.292 & 5.113 \\
0.1& 0 & 1.969 & 2.301 & 12.218 & 5.182 \\
0 & pn & 2.131 & 2.492 & 11.623 & 5.986 \\
\hline
\end{tabular}
\label{star}
\end{table}

First let us discuss the effect of the symmetry energy and the hyperon
couplings on the mass/radius curve. A larger $\Lambda_v$ gives rise to a
softer EOS and, therefore, a smaller radius. It is seen that when going from
$\Lambda_v= 0$ to 0.1 the radius of  a star with a mass $M=1-1.5\, M_\odot$ 
decreases by $\sim 0.3$ Km. This effect was already discussed within NLWM  for nucleonic stars 
\cite{Horowitz-01,fatto2010} and for hyperonic stars \cite{ rafael11}.
 In this last
paper it was shown that there exists a clear correlation between $L$ and star radius. However,
within the models discussed in \cite{rafael11}
the maximum mass did  not to depend  on $L$, while in the framework of
the QMC model there is a clear effect of almost 0.1$M_\odot$ if $\Lambda_v$
increases from 0 to 0.1. This is mainly due to the smaller strangeness
fraction inside the star.

The reduction of the attractiveness of $V_\Xi$ has a similar effect on the
maximum mass of the star, i.e., the mass increases $\sim 0.2 \, M_\odot$ if
$V_\Xi$ increases from -18 to 0 MeV.

We conclude that there is still quite a large uncertainty on the coupling of
hyperons to nuclear matter and therefore, there is still room for a very
massive star  such as the recently  measured pulsar J1614-2230 with a mass
$M=1.97 \pm 0.04$ \cite{demorest}, even including hyperons in the EOS.
This, however, is a particularly massive star.   Most of
the known pulsars \cite{chengmin} can be obtained by the present
models.

\section{Summary and Discussion}

We have proposed a modified QMC model which includes a nonlinear $\omega\rho$ 
coupling in the same fashion as it has been proposed for the NLMW 
\cite{Horowitz-01}. In QMC model the nucleons are
described as non-overlapping bags. The extra contribution allows the softening
of the symmetry energy at large densities. In NLWM or QMC without this
term the symmetry energy increases almost linearly with the baryonic density
giving rise to very hard stellar matter EOS. The inclusion of this term remedies
this problem and brings down the slope of the symmetry energy at saturation
density to values closer to the experimental predictions (see \cite{exp}
for a compilation of all constraints on $L$). 
With the modified QMC,  we may take advantage of
the already known good properties of the QMC together with a
symmetry energy that is not too hard.
We have also shown that the behavior of the modified QMC model is closer 
to the properties of nuclear relativistic models
with density dependent couplings such as TW, a relatvisitic nuclear model with
density dependent coulings \cite{tw}. Namely, the crust-core  transition
density is larger than the one predicted by the standard QMC and similar to 
TW and  the distillation effect in non homogeneous matter does not increase 
with density  as in NL3, but decreases as in TW.

We have also discussed the effect of the new EOS on the stellar properties.
Hyperons were included in the EOS, and, for the hyperon
couplings we took advantage of the fact that QMC predicts the hyperon
effective masses without being necessary to fix the hyperon-$\sigma$
couplings. We have used information from hypernuclei to fix the hyperon-$\omega$
coupling and the  hyperon-$\rho$ coupling was considered equal to the one of
the nucleon. Since there is a large uncertainty on hypernuclei with $\Sigma$
and $\Xi$, we have also tested the effect of increasing the potential $V_\Xi$
so that it becomes less attractive. It was shown that both the symmetry
energy and the
hyperon couplings have a strong effect on the mass and radius of the star. A
softer symmetry energy gives rise to smaller stars. Also the hyperon fraction
is affected: softer symmetry energy  corresponds to a smaller hyperon fraction
as already discussed in \cite{rafael11}. However, within QMC the density
dependence of the symmetry energy has also an effect on the maximum star mass
and this effect was not obtained in \cite{rafael11}.

It was also shown that the hyperon nuclear interaction defines the amount
of strangeness in the star, and, therefore, has a strong influence on the
maximum mass allowed. Even including hyperons in the QMC EOS we could explain
the mass of the pulsar  J1614-2230 if the cascade nuclear potential is set to
be very little attractive. More data on hypernuclei is needed to constrain the
hyperon-meson couplings.
\section*{Acknowledgments}
This work was partially supported by the Capes/FCT no. 232/09 bilateral
collaboration, by CNPq and FAPESC/1373/2010-0 (Brazil), by FCT and FEDER 
(Portugal) under the projects CERN/FP/83505/2008 and PTDC/FIS/113292/2009, 
and  by Compstar, an ESF Research Networking Programme.

\end{document}